\documentstyle[12pt]{article}
\begin{document}
{\pagestyle{empty}
\renewcommand{\thefootnote}{\fnsymbol{footnote}}
\rightline{TU-06/98}
\rightline{June 1998}
\rightline{~~~~~~~~~}
\vskip 1cm
\centerline{\Large \bf Brans-Dicke theory with the cosmological constant}
\vskip 2mm
\centerline{\Large \bf from $M_4\times\mbox{\boldmath{$Z$}}_2$ geometry}
\vskip 1cm

\centerline{%
  {Kunihiko Uehara \footnote
                   {Permanent e-mail address: uehara@tezukayama-u.ac.jp}}}
\vskip 8mm

\centerline{\it Department of Physics, Tezukayama University} 
\centerline{\it Tezukayama 631-8501, Japan}
\vskip 1mm
\centerline{and}
\vskip 1mm
\centerline{\it Abdus Salam International Centre for Theoretical Physics}
\centerline{\it P.\,O.\,Box 568, 34100 Trieste, Italy}

\vskip 15mm

\centerline{\bf Abstract}
\vskip 5mm

  The theory on $M_4\times\mbox{\boldmath{$Z$}}_2$ geometry
is applied to the Einstein gravity to yield the Brans-Dicke theory
on $M_4$ geometry.
The geometrical meaning and the relation between the curvatures
and the torsions are clarified.
The cosmological constant is also introduced into the pure Einstein action
on $M_4\times\mbox{\boldmath{$Z$}}_2$ in order to determine the explicit 
form of the cosmological term in the Brans-Dicke theory on $M_4$ geometry.

\vskip 0.4cm\noindent
PACS numbers: 02.40.-k, 04.50.+h, 46.10.+z, 98.80.-k
\hfil
\vfill
\newpage}
\setlength{\unitlength}{1mm}
\newcommand{\iDelta}{{\mit\Delta}}
\newcommand{\iGamma}{{\mit\Gamma}}
\indent
  The noncommutative geometry(NCG) of Connes\cite{Connes} has been successful
in producing geometrical interpretations of many grand unification models.
In these circumstances we have been giving geometrical meanings to them using 
the gauge theories\cite{SYM} on $M_4\times\mbox{\boldmath{$Z$}}_N$ geometry
without recourse to NCG, where the $\mbox{\boldmath{$Z$}}_N$ is 
the supplemented extra discrete space.
These approaches appear to be geometrically much simpler and clearer than NCG.

  Recently we have shown\cite{BDtheory} that the pure Einstein action on
$M_4\times\mbox{\boldmath{$Z$}}_2$ geometry exactly leads to the 
Brans-Dicke theory\cite{Brans} in four dimensional space-time $M_4$,
where we have used the equivalence assumption\cite{EAtheory} previously 
developed by three collaborators.
We have clarified the geometrical meaning of Riemann curvature tensors
which are classified to three kinds on this $M_4\times\mbox{\boldmath{$Z$}}_2$
manifold. 
Just after that our collaborators have written a paper\cite{STtheory} faithful 
to the fundamental concept of covariant differences.
But one may still ask why the new isometric condition in Ref.\cite{STtheory} 
would change the form of the curvature given by connections in Ref.\cite{BDtheory}, 
because the concept of the curvature given by the parallel transport 
has nothing to do with that of the distance which is introduced by 
the metric and the isometric condition above is given by the metric.

  It is the main purpose of this note to clarify the interrogative point 
above together with the definition of the curvature. 
And we also introduce the cosmological constant into the pure Einstein action 
on $M_4\times\mbox{\boldmath{$Z$}}_2$ geometry in order to determine the 
explicit form of the Brans-Dicke theory with the cosmological term on 
$M_4$ geometry in the Jordan frame, where the non-minimal coupling exists.
Hereafter we mainly follow the notations in Ref.\cite{STtheory}.
\vskip 5mm
  To begin with we introduce the tangent space $T(x,g)$ whose origin is $O(x,g)$
at a point $(x^\mu,g)$ on $M_4\times\mbox{\boldmath{$Z$}}_2$,
where $x^\mu\in M_4$ and $g=\{e{\rm(unit\ element)},r\}\in\mbox{\boldmath{$Z$}}_2$.
We consider a mapping of the origin $O(x+\iDelta x,g)$ onto $T(x,g)$
from $T(x+\iDelta x,g)$ which is another tangent space at a point
$(x^\mu+\iDelta x^\mu,g)$ neighboring to $(x^\mu,g)$.
The mapped point is denoted by a notation 
$U(x,x+\iDelta x,g)O(x+\iDelta x,g)$.
A covariant difference between $U(x,x+\iDelta x,g)O(x+\iDelta x,g)$ 
and $O(x,g)$ defines a vector $\mbox{\boldmath{$e$}}_{\mu }(x,g)$ 
on $T(x,g)$
\begin{eqnarray}
\triangle_x O(x,g)
&=& U(x,x+\iDelta x,g)O(x+\iDelta x,g)-O(x,g) \nonumber\\
&=& \mbox{\boldmath{$e$}}_\mu(x,g)\iDelta x^\mu.
\label{e201}
\end{eqnarray}

\noindent
Another vector $\mbox{\boldmath{$e$}}_r(x,g)$ can be got by the
covariant difference between the mapped point $V(x,g,g+r)O(x,g+r)$
and $O(x,g)$ on $T(x,g)$ in the same way above as follows:
\begin{eqnarray}
\triangle_r O(x,g)
&=& V(x,g,g+r)O(x,g+r)-O(x,g) \nonumber\\
&=& \mbox{\boldmath{$e$}}_r(x,g)\iDelta^r z(g),
\label{e202}
\end{eqnarray}

\noindent
where $z(g)$ is the coordinate corresponding to $g$ and 
\begin{equation}
\iDelta^r z(g)=z(g+r)-z(g)
\label{e203}
\end{equation}

\noindent
is proportional to the limiting process parameter\cite{EAtheory} which tends 
to zero where we assume the physics on two $M_4$ sheets should be equivalent 
to each other.
A set of vectors
\begin{equation}
\mbox{\boldmath{$e$}}_N(x,g)
=\left\{\mbox{\boldmath{$e$}}_\mu(x,g), \mbox{\boldmath{$e$}}_r(x,g)\,|\, 
        g=(e, r)\in \mbox{\boldmath{$Z$}}_2\right\}
\label{e204}
\end{equation}

\noindent
forms a basis on $T(x,g)$.
The parallel transport of the basis $\mbox{\boldmath{$e$}}_N(x+\iDelta x,g)$ from 
$T(x+\iDelta x,g)$ onto $T(x,g)$, which is given by a rotation 
$H^M_{\ N}(x,x+\iDelta x,g)$ of $\mbox{\boldmath{$e$}}_{M}(x,g)$, generates 
the parallel transported basis $\mbox{\boldmath{$e$}}_N^{\ H}(x+\iDelta x,g)$ as
\begin{equation}
\mbox{\boldmath{$e$}}_N^{\ H}(x+\iDelta x,g)
= \mbox{\boldmath{$e$}}_M(x,g)H^M_{\ N}(x,x+\iDelta x,g).
\label{e205}
\end{equation}

\noindent
The affine connection $\widehat{\iGamma}^M_{\,\,N\mu}(x,g)$ is defined
by
\begin{equation}
H^M_{\ N}(x,x+\iDelta x,g)
= \delta^M_{\ N} + \widehat{\iGamma}^M_{\,\,N\mu}(x,g)\iDelta x^\mu + O(\iDelta x^2).
\label{e206}
\end{equation}

\noindent
The covariant difference of $\mbox{\boldmath{$e$}}_{N}(x,g)$ can be defined on $M_{4}$ 
by use of two equations above, (\ref{e205}) and (\ref{e206}), as follows:
\begin{equation}
\triangle_x\mbox{\boldmath{$e$}}_N(x,g)
\equiv \mbox{\boldmath{$e$}}_N^{\ H}(x+\iDelta x,g) - \mbox{\boldmath{$e$}}_N(x,g)
= \mbox{\boldmath{$e$}}_M(x,g)\widehat{\iGamma}^M_{\,\,N\mu}(x,g)\iDelta x^\mu
+ O(\iDelta x^2).
\label{e207}
\end{equation}

\noindent
In the same way the covariant difference of $\mbox{\boldmath{$e$}}_{N}$ along 
$\mbox{\boldmath{$Z$}}_2$ is defined by
\begin{equation}
\triangle_r\mbox{\boldmath{$e$}}_N(x,g)
\equiv \mbox{\boldmath{$e$}}_N^{\ H}(x,g+r) - \mbox{\boldmath{$e$}}_N(x,g)
= \mbox{\boldmath{$e$}}_M(x,g)\widehat{\iGamma}^M_{\,\,Nr}(x,g)\iDelta^r z(g).
\label{e208}
\end{equation}

\noindent
The reason why the covariant difference in Eq.(\ref{e208}) is not associated 
with a term of $O(\iDelta^r z^2)$ is that the rotation matrix along 
$\mbox{\boldmath{$Z$}}_2$ has the form 
$H^M_{\,\,N}(x,g,g+r)=\delta^M_{\,\,N}+\widehat{\iGamma}^M_{\,\,Nr}(x,g)\iDelta^r z(g)$
due to the discreteness of $\mbox{\boldmath{$Z$}}_2$.
\vskip 5mm
  There are three kinds of torsion tensors on $M_4\times\mbox{\boldmath{$Z$}}_2$
geometry corresponding to the three path differences depicted in the Figure.
They are defined by
{\setcounter{enumi}{\value{equation}}
\addtocounter{enumi}{1}
\setcounter{equation}{0}
\renewcommand{\theequation}{\theenumi\alph{equation}}
\begin{eqnarray}
\left[\triangle_{1x},\triangle_{2x}\right]O(x,g)
&=& \mbox{\boldmath{$e$}}_M(x,g)\widehat{T}^M_{\,\,\mu\nu}(x,g)
          \iDelta_1 x^\mu\iDelta_2 x^\nu,
\label{e301a}\\
\left[\triangle_x,\triangle_r\right]O(x,g)
&=& \mbox{\boldmath{$e$}}_M(x,g)\widehat{T}^M_{\,\,\mu r}(x,g)
          \iDelta x^\mu\iDelta^r z(g),
\label{e301b}\\
\left[\triangle_r\triangle_r+2\triangle_r\right]O(x,g)
&=& \mbox{\boldmath{$e$}}_M(x,g)\widehat{T}^M_{\,\,rr}(x,g)
          \iDelta^r z(g)\iDelta^r z(g).
\label{e301c}
\end{eqnarray}
\setcounter{equation}{\value{enumi}}
}

\newpage
\setlength{\unitlength}{1mm}
\begin{picture}(75,35)(0,0)
 \put(30,5){\vector(-1,1){6}}\put(24,11){\line(-1,1){4}}
 \put(20,15){\vector(1,1){6}}\put(26,21){\line(1,1){4}}
 \put(30,5){\vector(1,1){6}}\put(36,11){\line(1,1){4}}
 \put(40,15){\vector(-1,1){6}}\put(34,21){\line(-1,1){4}}
 \put(30,5){\circle*{1}}
 \put(30,25){\circle*{1}}
 \put(29,27){$x$}
 \put(4,14){$x+\iDelta_1 x$}
 \put(41,14){$x+\iDelta_2 x$}
 \put(17,1){$x+\iDelta_1 x+\iDelta_2 x$}
 \put(22,14){$C_1$}
 \put(33,14){$C_2$}
 \put(25.5,-6){\bf\ (a)}
\end{picture}

\begin{picture}(0,0)(-53,-5)
 \put(10,5){\line(1,0){10}}
 \put(10,20){\line(1,0){10}}
 \put(35,5){\line(1,0){10}}
 \put(35,20){\line(1,0){10}}
 \put(20,5){\vector(0,1){8}}\put(20,13){\line(0,1){7}}
 \put(20,20){\vector(1,0){8}}\put(28,20){\line(1,0){7}}
 \put(20,5){\vector(1,0){8}}\put(28,5){\line(1,0){7}}
 \put(35,5){\vector(0,1){8}}\put(35,13){\line(0,1){7}}
 \put(20,5){\circle*{1}}
 \put(35,20){\circle*{1}}
 \put(13,1){$x+\iDelta x$}
 \put(34,1){$x$}
 \put(13,22){$x+\iDelta x$}
 \put(34,22){$x$}
 \put(49,4){$g+r$}
 \put(49,19){$g$}
 \put(21,16){$C_3$}
 \put(30,6){$C_4$}
 \put(23,-6){\bf\ (b)}
\end{picture} 

\begin{picture}(0,0)(-100,-10)
 \put(25,5){\line(1,0){10}}
 \put(25,20){\line(1,0){10}}
 \put(35,5){\line(1,0){10}}
 \put(35,20){\line(1,0){10}}
 \put(35,20){\vector(0,-1){8}}\put(35,12){\line(0,-1){6.5}}
 \put(36,5.5){\vector(0,1){7.5}}\put(36,13){\line(0,1){7}}
 \put(35.5,5.5){\oval(1,1)[b]} 
 \put(35.5,20){\circle*{1}}
 \put(30,1){$(x,g+r)$}
 \put(30,22){$(x,g)$}
 \put(31.5,-7){\bf\ (c)}
\end{picture}

\noindent
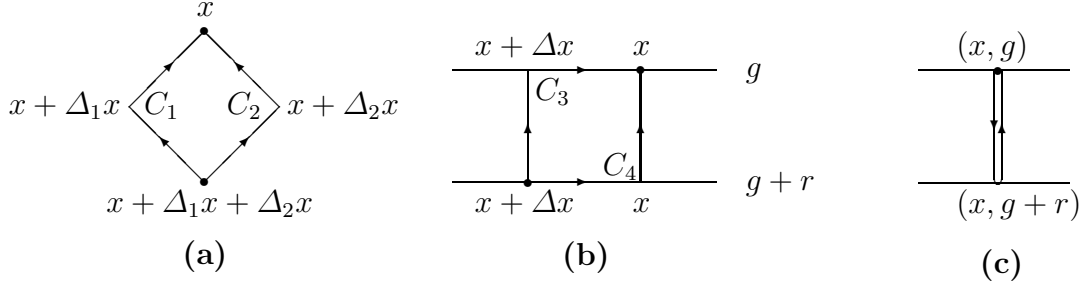
{\small{\bf Figure.} Path differences where parallel transports are compared
for (a)\,the first, (b)\,the second and (c)\,the third kind.}

\vskip 8mm

The torsion of the first kind $\widehat{T}^M_{\,\,\mu\nu}$ is the conventional one,
which is derived by comparing with two mappings of the origin 
$O(x+\iDelta_1 x+\iDelta_2 x,g)$ from $T(x+\iDelta_1 x+\iDelta_2 x,g)$ onto $T(x,g)$ 
along two paths $C_1$ and $C_2$ depicted in Fig.(a).
They are given by
\begin{eqnarray}
C_1
&=&U(x,x+\iDelta_1 x,g)U(x+\iDelta_1 x,x+\iDelta_1 x+\iDelta_2 x,g)
   O(x+\iDelta_1 x+\iDelta_2 x,g),
\label{e302}\\
C_2
&=&U(x,x+\iDelta_2 x,g)U(x+\iDelta_2 x,x+\iDelta_1 x+\iDelta_2 x,g)
   O(x+\iDelta_1 x+\iDelta_2 x,g).
\label{e303}
\end{eqnarray}

\noindent
The straightforward calculation of the difference between $C_1$ and $C_2$ gives 
the torsion of the first kind as follows:
\begin{eqnarray}
C_1-C_2=(\triangle_{1x}\triangle_{2x}-\triangle_{2x}\triangle_{1x})O(x,g),
\label{e304}
\end{eqnarray}

\noindent
which is just Eq.(\ref{e301a}).
The torsion of the second kind $\widehat{T}^M_{\,\,\mu r}$ is obtained in the 
same way above by considering two mappings of the origin $O(x+\iDelta x,g+r)$ 
from $T(x+\iDelta x,g+r)$ onto $T(x,g)$ along two paths $C_3$ and $C_4$
depicted in Fig.(b).
They are given by
\begin{eqnarray}
C_3 &=& U(x,x+\iDelta x,g)V(x+\iDelta x,g,g+r)O(x+\iDelta x,g+r),
\label{e305}\\
C_4 &=& V(x,g,g+r)U(x,x+\iDelta x,g+r)O(x+\iDelta x,g+r).
\label{e306}
\end{eqnarray}

\noindent
The difference between $C_3$ and $C_4$ gives the torsion of the second kind 
by the similar calculation above as
\begin{eqnarray}
C_3-C_4=(\triangle_x\triangle_r - \triangle_r\triangle_x)O(x,g),
\label{e307}
\end{eqnarray}

\noindent
which is identical to Eq.(\ref{e301b}).
Finally the torsion of the third kind $\widehat{T}^M_{\,\,rr}$ is derived considering the
mappings of the origin $O(x,g)$ from $T(x,g)$ onto $T(x,g+r)$ and again onto
the same $T(x,g)$ depicted in Fig.(c).
The difference between the mapped point and $O(x,g)$ gives the torsion
after some calculations\cite{STtheory} as
\begin{equation}
V(x,g,g+r)V(x,g+r,g)O(x,g)-O(x,g)
=\left(\triangle_r\triangle_r+2\triangle_r\right)O(x,g).
\label{e308}
\end{equation}

\noindent
By using the relation $\iDelta^r z(g+r)=-\iDelta^r z(g)$ and 
the Eqs.(\ref{e201}), (\ref{e202}), (\ref{e207}) and (\ref{e208}), 
one will find the following three kind of torsions in terms of connections:
{\setcounter{enumi}{\value{equation}}
\addtocounter{enumi}{1}
\setcounter{equation}{0}
\renewcommand{\theequation}{\theenumi\alph{equation}}
\begin{eqnarray}
\widehat{T}^M_{\,\,\mu\nu}(x,g)
&=& \widehat{\iGamma}^M_{\,\,\nu\mu}(x,g)
   -\widehat{\iGamma}^M_{\,\,\mu\nu}(x,g),
\label{e309a}\\
\widehat{T}^M_{\,\,\mu r}(x,g)
&=& \widehat{\iGamma}^M_{\,\,r\mu}(x,g)
   -\widehat{\iGamma}^M_{\,\,\mu r}(x,g),
\label{e309b}\\
\widehat{T}^M_{\,\,rr}(x,g)
&=&-\widehat{\iGamma}^M_{\,\,rr}(x,g).
\label{e309c}
\end{eqnarray}
\setcounter{equation}{\value{enumi}}
}\noindent\hskip -.5em
The torsions of the first and second kind vanish when the affine connection
$\widehat{\iGamma}^L_{\,\,MN}(x,g)$ is symmetric with respect to $M$ and $N$, namely,
\begin{equation}
\widehat{\iGamma}^L_{\,\,MN}(x,g)=\widehat{\iGamma}^L_{\,\,NM}(x,g),
\label{e310}
\end{equation}

\noindent
nevertheless the torsion of the third kind $\widehat{T}^M_{\,\,rr}$ generally remains finite.
\vskip 5mm
  There are also three kinds of curvature tensors corresponding to three kinds of 
path differences depicted in the Figure.
They are defined by
{\setcounter{enumi}{\value{equation}}
\addtocounter{enumi}{1}
\setcounter{equation}{0}
\renewcommand{\theequation}{\theenumi\alph{equation}}
\begin{eqnarray}
\left[\triangle_{1x},\triangle_{2x}\right]\mbox{\boldmath{$e$}}_N
&=& \mbox{\boldmath{$e$}}_M\widehat{\mbox{\boldmath{$R$}}}^M_{\ N\mu\nu}
    \iDelta_1 x^\mu\iDelta_2 x^\nu
   +\mbox{\boldmath{$e$}}_M\widehat{\iGamma}^M_{\,\,NL}
    \widehat{T}^L_{\mu\nu}\iDelta_1 x^\mu\iDelta_2 x^\nu,
\label{e401a}\\
\left[\triangle_x,\triangle_r\right]\mbox{\boldmath{$e$}}_N
&=& \mbox{\boldmath{$e$}}_M\widehat{\mbox{\boldmath{$R$}}}^M_{\ N\mu r}
    \iDelta x^\mu\iDelta^r z
   +\mbox{\boldmath{$e$}}_M\widehat{\iGamma}^M_{\,\,NL}
    \widehat{T}^L_{\mu r}\iDelta x^\mu\iDelta^r z,
\label{e401b}\\
\left[\triangle_r\triangle_r + 2\triangle_r\right]\mbox{\boldmath{$e$}}_N
&=& \mbox{\boldmath{$e$}}_M\widehat{\mbox{\boldmath{$R$}}}^M_{\ Nrr}
    \iDelta^r z\iDelta^r z
   +\mbox{\boldmath{$e$}}_M\widehat{\iGamma}^M_{\,\,NL}
    \widehat{T}^L_{rr}\iDelta^r z\iDelta^r z.
\label{e401c}
\end{eqnarray}
\setcounter{equation}{\value{enumi}}
}\noindent\hskip -.4em
As far as we constrain ourselves in the theory where $\widehat{\iGamma}^L_{\,\,MN}$ is
symmetric with respect to $M$ and $N$, Eqs.(\ref{e401a}) and (\ref{e401b}) reproduce
the same definitions in Ref.\cite{STtheory}.  But even in this case the curvature of the 
third kind in Eq.(\ref{e401c}) appears different due to the existence of the
torsion term as follows together with curvatures of the first and second kind:
{\setcounter{enumi}{\value{equation}}
\addtocounter{enumi}{1}
\setcounter{equation}{0}
\renewcommand{\theequation}{\theenumi\alph{equation}}
\begin{eqnarray}
\widehat{\mbox{\boldmath{$R$}}}^M_{\ N\mu\nu}
&=& \partial_\mu\widehat{\iGamma}^M_{\,\,N\nu}-\partial_\nu\widehat{\iGamma}^M_{\,\,N\mu}
   +\widehat{\iGamma}^M_{\,\,L\mu}\widehat{\iGamma}^L_{\,\,N\nu}
   -\widehat{\iGamma}^M_{\,\,L\nu}\widehat{\iGamma}^L_{\,\,N\mu},
\label{e402a}\\
\widehat{\mbox{\boldmath{$R$}}}^M_{\ N\mu r}
&=& \partial_\mu\widehat{\iGamma}^M_{\,\,Nr}-\partial_r\widehat{\iGamma}^M_{\,\,N\mu}
   +\widehat{\iGamma}^M_{\,\,L\mu}\widehat{\iGamma}^L_{\,\,Nr}
   -\widehat{\iGamma}^M_{\,\,Lr}\widehat{\iGamma}^L_{\,\,N\mu},
\label{e402b}\\
\widehat{\mbox{\boldmath{$R$}}}^M_{\ Nrr}
&=& -\partial_r\widehat{\iGamma}^M_{\,\,Nr}
    -\widehat{\iGamma}^M_{\,\,Lr}\widehat{\iGamma }^L_{\,\,Nr}
    +\widehat{\iGamma}^M_{\,\,NL}\widehat{\iGamma }^L_{\,\,rr}.
\label{e402c}
\end{eqnarray}
\setcounter{equation}{\value{enumi}}
}\noindent\hskip -.3em
In order to evaluate these curvatures in the concept of distance,
the metric must be introduced onto the manifold.

\vskip 5mm

  The manifold $M_{4}\times \mbox{\boldmath{$Z$}}_2$ can be regarded as 
the five dimensional Kaluza-Klein space except the fifth dimension is replaced 
by two points $z(e)$ and $z(r)$.
The line element $\iDelta s$ of this space is assumed here as 
\begin{eqnarray}
\iDelta s^2
&=& g_{\mu\nu}(x)\iDelta x^\mu\iDelta x^\nu+\lambda^2(x)\iDelta z^2 \nonumber\\
&=& G_{MN}(x)\iDelta x^M\iDelta x^N,  
\label{e403}
\end{eqnarray}

\noindent
where
\begin{equation}
\iDelta x^N
= (\iDelta x^\mu,\iDelta x^r\equiv\iDelta z=z(r)-z(e))
\label{e404}
\end{equation}

\noindent
and $G_{MN}(x)$ is the five dimensional metric on $M_{4}\times\mbox{\boldmath{$Z$}}_2$,
which is defined by the inner product of basis vectors 
\begin{equation}
G_{MN}(x)=\mbox{\boldmath{$e$}}_M(x,g)\cdot\mbox{\boldmath{$e$}}_N(x,g).
\label{e405}
\end{equation}

\noindent
Let the manifold $M_4\times\mbox{\boldmath{$Z$}}_2$ be isometric, namely,
any inner product of vectors is invariant under the parallel transport as
\begin{equation}
G_{MN}(x+\iDelta x)
= \mbox{\boldmath{$e$}}_M(x+\iDelta x,g)\cdot\mbox{\boldmath{$e$}}_N(x+\iDelta x,g)
= \mbox{\boldmath{$e$}}_M^{\ H}(x+\iDelta x,g)
  \cdot\mbox{\boldmath{$e$}}_N^{\ H}(x+\iDelta x,g).
\label{e406}
\end{equation}

\noindent
Substituting Eqs.(\ref{e207}) and (\ref{e208}) into (\ref{e406}),
the relations between the metric and the connections can be obtained for 
$M_4$ and $\mbox{\boldmath{$Z$}}_2$, respectively
\begin{eqnarray}
\partial_\lambda G_{MN}
&=& \widehat{\iGamma}_{M\lambda N}+\widehat{\iGamma}_{N\lambda M},
\label{e407}\\
\partial_r G_{MN}
&=& \widehat{\iGamma}_{MrN}+\widehat{\iGamma}_{NrM}
   +\widehat{\iGamma}_{KrM}\widehat{\iGamma }^K_{\,\,rN}\iDelta^r z,
\label{e408}
\end{eqnarray}

\noindent
where $\widehat{\iGamma}_{MLN}\equiv G_{MK}\widehat{\iGamma}^K_{\,\,LN}$.
The limit $\iDelta^r z\rightarrow 0$ in Eqs.(\ref{e407}) and (\ref{e408}) yields
\begin{eqnarray}
\partial_{\lambda}G_{MN}
&=& \iGamma_{M\lambda N}+\iGamma_{N\lambda M},
\label{e409}\\
\partial_r G_{MN}
&=& \iGamma_{MrN}+\iGamma_{NrM},
\label{e410}
\end{eqnarray}

\noindent
where notations without a hat mean that these quantities are 
independent of $g$.
One can obtain the well-known expression of $\iGamma_{LMN}$ in terms of 
$G_{MN}$ using Eqs.(\ref{e310}), (\ref{e409}) and (\ref{e410}) 
\begin{equation}
\iGamma_{LMN}
=\frac{1}{2}(\partial_M G_{LN}+\partial_N G_{LM}-\partial_L G_{MN}).
\label{e411}
\end{equation}

\noindent
Since the metric $G_{MN}$ is given by (\ref{e403}), one immediately get followings: 
\begin{eqnarray}
\iGamma_{\lambda\mu\nu}
&=& \frac{1}{2}(\partial_\mu g_{\lambda\nu}+\partial_{\nu}g_{\lambda\mu}
                                -\partial_\lambda g_{\mu\nu}),
\label{e412}\\
\iGamma_{r\mu\nu}
&=& \iGamma_{\mu\nu r}
 =  \iGamma_{\mu r\nu}
 =  0,
\label{e413}\\
\iGamma_{rr\mu}
&=& \iGamma_{r\mu r}
 = -\iGamma_{\mu rr}
 =  \lambda\partial_\mu\lambda,
\label{e414}\\
\iGamma_{rrr} &=& 0.
\label{e415}
\end{eqnarray}

On the other hand, in the same limit $\iDelta^r z\rightarrow 0$ after 
differentiating the Eq.(\ref{e408}) with respect to $z(g)$ together with 
the fact $\partial_r\iDelta^r z(g)=-2$, one can get 
\begin{equation}
(\partial_r\widehat{\iGamma}_{MrN}+\partial_r\widehat{\iGamma}_{NrM})_0
= 2\iGamma_{KrM}\iGamma^K_{\,\,rN},
\label{e416}
\end{equation}

\noindent
where the quantity with a subscript $0$ also means that quantity is independent of $g$.
By using Eqs.(\ref{e413}), (\ref{e414}) and (\ref{e416}) with $M=\mu$ and $N=\nu$, 
one can see
\begin{equation}
(\partial_r\widehat{\iGamma}_{\mu r\nu}+\partial_r\widehat{\iGamma}_{\nu r\mu})_0
= 2\iGamma_{Kr\mu}\iGamma^K_{\,\,r\nu}
= 2\iGamma_{rr\mu}\iGamma^r_{\,\,r\nu}
\label{e417}
\end{equation}

\noindent
and by Eq.(\ref{e407}) 
\begin{equation}
\partial_\nu G_{\mu r}
=\widehat{\iGamma}_{\mu\nu r}+\widehat{\iGamma}_{r\nu\mu}=0,
\label{e418}
\end{equation}

\noindent
which with Eq.(\ref{e310}) reads
\begin{equation}
\widehat{\iGamma}_{\mu\nu r}
=-\widehat{\iGamma}_{r\nu\mu}
=-\widehat{\iGamma}_{r\mu\nu}
= \widehat{\iGamma}_{\nu\mu r}
\label{e419}
\end{equation}

\noindent
From this equation (\ref{e419}) one can see that the first or the second term
in the left-hand side of (\ref{e417}) is equal to each other, hence  
\begin{equation}
(\partial_r\widehat{\iGamma}_{\mu\nu r})_0
= \iGamma_{rr\mu}\iGamma^r_{\,\,r\nu}
= -(\partial_r\widehat{\iGamma}_{r\mu\nu})_0.
\label{e420}
\end{equation}

\noindent
Another important equation to evaluate the curvatures comes from 
Eq.(\ref{e416}) with $M=N=r$
\begin{equation}
(\partial_r\widehat{\iGamma}_{rrr})_0
=\iGamma_{Krr}\iGamma^K_{\,\,rr}
=\iGamma_{\rho rr}\iGamma^\rho_{\,\,rr}.
\label{e421}
\end{equation}
\vskip 5mm
  In the limit $\Delta^r z\rightarrow 0$, three curvature tensors 
(\ref{e402a})-(\ref{e402c}) can be rewritten using 
Eqs.(\ref{e409}) and (\ref{e410}) as follows: 
{\setcounter{enumi}{\value{equation}}
\addtocounter{enumi}{1}
\setcounter{equation}{0}
\renewcommand{\theequation}{\theenumi\alph{equation}}
\begin{eqnarray}
\mbox{\boldmath{$R$}}_{MN\mu\nu}
&=& \partial_\mu\iGamma_{MN\nu}-\partial_\nu\iGamma_{MN\mu}
   -\iGamma_{LM\mu}\iGamma^L_{\,\,N\nu}+\iGamma_{LM\nu}\iGamma^L_{\,\,N\mu},
\label{e501a}\\
\mbox{\boldmath{$R$}}_{MN\mu r}
&=& \partial_\mu\iGamma_{MNr}-(\partial_r\widehat{\iGamma}_{MN\mu})_0
   -\iGamma_{LM\mu}\iGamma^L_{\,\,Nr}+\iGamma_{LMr}\iGamma^L_{\,\,N\mu},
\label{e501b}\\
\mbox{\boldmath{$R$}}_{MNrr}
&=& -(\partial_r\widehat{\iGamma}_{MNr})_0+\iGamma_{LMr}\iGamma^L_{\,\,Nr}
    +\iGamma_{MNL}\iGamma^L_{\,\,rr}.
\label{e501c}
\end{eqnarray}
\setcounter{equation}{\value{enumi}}
}
\noindent\hskip -.5em
The equation (\ref{e501a}) with $M=\rho$ and $N=\nu$ reads
\begin{eqnarray}
\mbox{\boldmath{$R$}}_{\rho\sigma\mu\nu}
&=& \partial_\mu\iGamma_{\rho\sigma\nu}-\partial_\nu\iGamma_{\rho\sigma\mu}
   -\iGamma_{L\rho\mu}\iGamma^L_{\,\,\sigma\nu}
   +\iGamma_{L\rho\nu}\iGamma^L_{\,\,\sigma\mu} \nonumber\\
&=& \partial_\mu\iGamma_{\rho\sigma\nu}-\partial_\nu\iGamma_{\rho\sigma\mu}
   -\iGamma_{\lambda\rho\mu}\iGamma^\lambda_{\,\,\sigma\nu}
   +\iGamma_{\lambda\rho\nu}\iGamma^\lambda_{\,\,\sigma\mu}
 =  R_{\rho\sigma\mu\nu},
\label{e502}
\end{eqnarray}

\noindent
which gives the conventional scalar curvature on $M_4$
\begin{equation}
R=g^{\mu\nu}R^\rho_{\ \mu\rho\nu}.
\label{e503}
\end{equation}

\noindent
The equation (\ref{e501b}) together with (\ref{e413}), (\ref{e414}) and 
(\ref{e420}) gives the relevant component $\mbox{\boldmath{$R$}}_{\nu r\mu r}$ as 
\begin{eqnarray}
\mbox{\boldmath{$R$}}_{\nu r\mu r}
&=& \partial_\mu\iGamma_{\nu rr}-(\partial_r\widehat{\iGamma}_{\nu r\mu})_0
   -\iGamma_{L\nu\mu}\iGamma^L_{\,\,rr}
   +\iGamma_{L\nu r}\iGamma^L_{\,\,r\mu} \nonumber\\
&=& \partial_\mu\iGamma_{\nu rr}-\iGamma_{rr\nu}\iGamma^r_{\,\,r\mu}
   -\iGamma_{\rho\nu\mu}\iGamma^\rho_{\,\,rr}
   +\iGamma_{r\nu r}\iGamma^r_{\,\,r\mu} \nonumber\\
&=&-\nabla_{\!\mu}(\lambda\partial_\nu\lambda),
\label{e504}
\end{eqnarray}

\noindent
where $\nabla_{\!\mu}$ is the covariant derivative in $M_4$, hence, one get
\begin{equation}
\mbox{\boldmath{$R$}}^\rho_{\ r\rho r}
= -\nabla^\rho(\lambda\partial_\rho\lambda).
\label{e505}
\end{equation}

\noindent
In the same way one can get 
\begin{eqnarray}
\mbox{\boldmath{$R$}}_{r\nu r\mu}
&=& (\partial_r\widehat{\iGamma}_{r\nu\mu})_0-\partial_\mu\iGamma_{r\nu r}
   -\iGamma_{Lrr}\iGamma^L_{\,\,\nu\mu}
   +\iGamma_{Lr\mu}\iGamma^L_{\,\,\nu r} \nonumber\\
&=&-\iGamma_{rr\nu}\iGamma^r_{\,\,r\mu}-\partial_\mu\iGamma_{r\nu r}
   -\iGamma_{\rho rr}\iGamma^\rho_{\,\,\nu\mu}
   +\iGamma_{rr\mu}\iGamma^r_{\,\,\nu r} \nonumber\\
&=&-\nabla_{\!\mu}(\lambda\partial_\nu\lambda),
\label{e506}
\end{eqnarray}

\noindent
so that 
\begin{equation}
\mbox{\boldmath{$R$}}^r_{\ \mu r\nu}
= G^{rr}\mbox{\boldmath{$R$}}_{r\mu r\nu}
=-\lambda^{-2}\nabla_{\!\nu}(\lambda\partial_\mu\lambda).
\label{e507}
\end{equation}

\noindent
The equation (\ref{e501c}) with $M=N=r$ together with (\ref{e414}) and 
(\ref{e421}) gives
\begin{eqnarray}
\mbox{\boldmath{$R$}}_{rrrr}
&=&-(\partial_r\widehat{\iGamma}_{rrr})_0
   +\iGamma_{Lrr}\iGamma^L_{\,\,rr}
   +\iGamma_{rrL}\iGamma^L_{\,\,rr} \nonumber\\
&=&-(\partial_r\widehat{\iGamma}_{rrr})_0
   +\iGamma_{\rho rr}\iGamma^\rho_{\,\,rr}
   +\iGamma_{rr\rho}\iGamma^\rho_{\,\,rr} \nonumber\\
&=&-\lambda^2\partial_\rho\lambda\partial^\rho\lambda,
\label{e508}
\end{eqnarray}

\noindent
hence 
\begin{equation}
\mbox{\boldmath{$R$}}^r_{\,\,rrr}
= G^{rr}\mbox{\boldmath{$R$}}_{rrrr}
=-\partial_\rho\lambda\partial^\rho\lambda,
\label{e509}
\end{equation}

\noindent
which is identical to that in the previous work\cite{BDtheory} as is expected.
The Ricci curvature on $M_4\times\mbox{\boldmath{$Z$}}_2$ is therefore given by
\begin{eqnarray}
\mbox{\boldmath{$R$}}_{MN}
&\equiv& \mbox{\boldmath{$R$}}^L_{\,\,MLN} \nonumber\\
&=&      \mbox{\boldmath{$R$}}^\rho_{\,\,M\rho N}
        +\mbox{\boldmath{$R$}}^r_{\,\,MrN},
\label{e510}
\end{eqnarray}

\noindent
hence
\begin{eqnarray}
\mbox{\boldmath{$R$}}_{\mu\nu}
&=& R_{\mu\nu}+\mbox{\boldmath{$R$}}^r_{\,\,\mu r\nu} \nonumber\\
&=& R_{\mu\nu}-\lambda^{-2}\nabla_{\!\nu}(\lambda\partial_\mu\lambda),
\label{e511}\\
\mbox{\boldmath{$R$}}_{rr}
&=& \mbox{\boldmath{$R$}}^\rho_{\,\,r\rho r}
   +\mbox{\boldmath{$R$}}^r_{\,\,rrr} \nonumber\\
&=&-\nabla_{\!\rho}(\lambda\partial^\rho\lambda)
   -\partial_\rho\lambda\partial^\rho\lambda,
\label{e512}
\end{eqnarray}

\noindent
where $R_{\mu\nu}\equiv\mbox{\boldmath{$R$}}^\rho_{\,\,\mu\rho\nu}$ is 
the conventional four dimensional Ricci curvature.
Finally, therefore, one get the scalar curvature on $M_4\times\mbox{\boldmath{$Z$}}_2$
\begin{eqnarray}
\mbox{\boldmath{$R$}}
&\equiv& G^{MN}\mbox{\boldmath{$R$}}_{MN} \nonumber\\
&=&      g^{\mu\nu}\mbox{\boldmath{$R$}}_{\mu\nu}
        +G^{rr}\mbox{\boldmath{$R$}}_{rr} \nonumber\\
&=&      R-2\lambda^{-2}\nabla_{\!\rho}(\lambda\partial^\rho\lambda)
        -\lambda^{-2}\partial_\rho\lambda\partial^\rho\lambda.
\label{e513}
\end{eqnarray}

\noindent
The action of the gravity can be now obtained by using the fact \ 
$\det(G_{MN})=\det(g_{\mu\nu})\lambda^2$
\begin{eqnarray}
I
&=&\int_{M_4}\int_{Z_2}\sqrt{-\det(G_{MN})}\,\mbox{\boldmath{$R$}} \nonumber\\
&=&\int_{M_4}\sqrt{-\det(g_{\mu\nu})}\,
   \lambda[R-2\lambda^{-2}\nabla_{\!\rho}(\lambda\partial^\rho\lambda)
          -\lambda^{-2}\partial_\rho\lambda\partial^\rho\lambda] \nonumber\\
&=&\int_{M_4}\sqrt{-\det(g_{\mu\nu})}\,
   [\lambda R-\frac{3}{\lambda}\partial_\rho\lambda\partial^\rho\lambda],
\label{e514}
\end{eqnarray}

\noindent
which is just the Brans-Dicke theory with Dicke constant $\omega=3$.
\vskip 5mm
  Inspired by the result above, we also introduce the cosmological constant 
into the Einstein gravity on $M_4\times\mbox{\boldmath{$Z$}}_2$ geometry 
in order to determine the explicit form of the cosmological term 
in the Brans-Dicke theory on $M_4$ geometry.
The calculation itself is as same as the pure Einstein case, and one obtains the
action of the Brans-Dicke theory with the cosmological term $\Lambda$ in 
the Jordan frame as
\begin{eqnarray}
I
&=&\int_{M_4}\int_{Z_2}\sqrt{-\det(G_{MN})}\,(\mbox{\boldmath{$R$}}-2\Lambda) \nonumber\\
&=&\int_{M_4}\sqrt{-\det(g_{\mu\nu})}\,
   [\lambda(R-2\Lambda)-\frac{3}{\lambda}\partial_\mu\lambda\partial^{\,\mu}\lambda],
\label{e601}
\end{eqnarray}

\noindent
to which we got the exact solutions\cite{Uehara} with an undetermined Dicke
constant for the spatially flat cosmology some years ago.
The choice of this form was done at that time by another reason, that is,
such a cosmological term simply leads to the mass term of the scalar field,
although the general form of the scalar-tensor gravitation theory\cite{Wagoner}
including the cosmological term was already proposed to attempt the general theory.
\vskip 5mm
  On the basis of the equivalence assumption and also of the isometric 
conditions (\ref{e407}) and (\ref{e408}), we have derived the Brans-Dicke theory 
from the manifold $M_4\times\mbox{\boldmath{$Z$}}_2$.
We have also clarified the geometrical relation between torsions and curvature 
in this space.
In the work\cite{STtheory} the contribution from the torsion to the curvature 
in Eq.(\ref{e402c}) has not been taken into account. 
The existence of the torsion term has actually played an essential role to reproduce 
the result given by the concept of the parallel transport.\cite{BDtheory}
As for the Dicke constant, one may get a greater value of $\omega$ if the
structure of the discrete space has more degrees of freedom.
Finally we have determined the explicit form of the cosmological term 
in the Brans-Dicke theory on $M_4$ geometry within the framework of the theory.
\vskip 5mm
\noindent
{\bf Acknowledgements}

  The author would like to express his gratitude for the hospitality at the
Abdus Salam International Centre for Theoretical Physics.
This work is supported partly by the research abroad grant of Tezukayama
university and partly by the promotion and mutual aid corporation for
private schools of Japan.
\vskip 5mm
\end{document}